# Helix alignment, chevrons, and edge dislocations in twist-bend ferroelectric nematics

Bijaya Basnet[1,2], Priyanka Kumari[1,2], Sathyanarayana Paladugu[1], Damian Pociecha[3], Jakub Karcz[4], Przemysław Kula[4], Nataša Vaupotič[5,6], Ewa Górecka[3], and Oleg D Lavrentovich [1,2,3,7*]

[1]*Advanced Materials and Liquid Crystal Institute, Kent State University, Kent, OH 44242, USA*
[2]*Materials Science Graduate Program, Kent State University, Kent, OH 44242, USA*
[3]*Faculty of Chemistry, University of Warsaw, Zwirki i Wigury 101, Warsaw 02-089, Poland*
[4]*Faculty of Advanced Technology and Chemistry, Military University of Technology, Warsaw, Poland*
[5]*Jozef Stefan Institute, Jamova 39, 1000 Ljubljana, Slovenia*
[6]*Department of Physics, Faculty of Natural Sciences and Mathematics, University of Maribor, Koroška 160, 2000 Maribor, Slovenia*
[7] *Department of Physics, Kent State University, Kent, OH 44242, USA*
*Authors for correspondence: e-mails: olavrent@kent.edu

**Abstract.** We explore surface alignment and edge dislocations in the recently discovered twist-bend ferroelectric nematic, $N_{TBF}$, in which the vector of spontaneous polarization follows an oblique helicoidal trajectory around a polar twist-bend axis. In a planar cell, the polar axis aligns at some angle to the rubbing direction to mitigate the surface electric charge. We demonstrate that the pseudolayers in planar cells form chevron defects, a hallmark defect of one-dimensionally positionally ordered phases, such as smectic A and smectic C. The polar character of the twist-bend axis prevents the cores of $N_{TBF}$ edge dislocations from splitting into semi-integer disclinations, in stark contrast to dislocations in paraelectric and ferroelectric chiral nematics. The tilt of pseudolayers around the defect core allows us to estimate the elastic penetration length as being close to the pitch of $N_{TBF}$. Compression/dilation stresses around the core modify the heliconical tilt angle of molecules as evidenced by a substantial variation in local birefringence. The climb of dislocations exhibits high mobility, allowing the system to equilibrate the temperature-dependent pitch. The uncovered properties facilitate the development of $N_{TBF}$ materials for electro-optical applications, such as electrically controlled diffraction lattices and structural colors.

I. INTRODUCTION

The recent identification of a ferroelectric nematic, $N_F$, liquid crystal (1-4) has been followed by the discoveries of many more ferroelectric mesomorphic phases, such as ferroelectric smectic A (5-10), antiferroelectric SmA (11), ferroelectric smectic C (10, 12), ferroelectric smectic C with heliconical structure (11) and, very recently, a polar version of the twist-bend nematic (13-15). This polar heliconical phase has been called a twist-bend ferroelectric nematic ($N_{TBF}$) by Karcz et al. (16), and a heliconical ferroelectric nematic abbreviated $^{HC}N_F$ by Nishikawa et al. (17). Since the two abbreviations refer to the same state of soft matter (17), we use the abbreviation $N_{TBF}$ in what follows. A similar oblique helicoidal structure, but with an additional smectic layering caused by a density wave along the twist-bend axis, has been discovered by Gibb et al.(11) and abbreviated $SmC_P^H$. The $N_{TBF}$ and $SmC_P^H$ are formed by achiral molecules. In both phases, the pitch is in the order of hundreds of nanometers, which is much larger than the 10 nm pitch of non-polar $N_{TB}$ phase (14, 15). The long pitch suggests that the ambidextrous chirality of these phases is rooted in polar interactions rather than in the bent shape of molecules, which is the case of $N_{TB}$ (18). Indeed, the $N_{TBF}$ forming molecules are straight rod- or slightly plank-like with a strong longitudinal dipole, approximately 13 D (16, 17). When an electric field is applied, the pitch and the conic angle of $N_{TBF}$ both decrease till the polarization aligns fully along the field (16, 17). The electric tunability of the pitch makes the $N_{TBF}$ similar to the electrically tunable oblique helicoidal cholesteric $Ch_{OH}$ (19, 20). Thanks to the submicron pitch, both the $N_{TBF}$ and $Ch_{OH}$ can be used for tunable selective reflection of light in a broad spectral range. Unlike the $Ch_{OH}$, which is sustainable only in the presence of an electric (19, 20) or magnetic (21) field, the newly discovered polar $N_{TBF}$ is stable in the absence of external fields.

The polarization field of a uniform unconstrained $N_{TBF}$ writes in the Cartesian coordinates as a polar analog of the twist-bend nematic $N_{TB}$ director field (14, 15, 18, 22, 23):

$$\mathbf{P} = \{P_x, P_y, P_z\} = P\{\sin\theta\cos\varphi,\ \sin\theta\sin\varphi, \cos\theta\}, \qquad (1)$$

where $\theta$ is the conical tilt angle of the local polarization with respect to the helicoidal axis $\mathbf{q}$ directed along the z-axis, $\varphi = qz$ is the heliconical phase, $q = 2\pi/\mathcal{P}$ and $\mathcal{P}$ is the pitch. Besides the structural similarity with the $N_{TB}$ and $Ch_{OH}$, the $N_{TBF}$ also connects to a paraelectric chiral nematic ($N^*$), also called a cholesteric (24), and a ferroelectric cholesteric ($N_F^*$) (25-28), in which $\theta = \pi/2$. An important distinction of the $N_{TBF}$ from the $N^*$, $N_F^*$ and $Ch_{OH}$ is that it bears both local polarity (a nonvanishing **P**) and global polarity, caused by the fact that a flip of the twist axis **q** by $\pi$ brings about a distinct state, i.e. $\mathbf{q} \neq -\mathbf{q}$.

In this study, we explore alignment and structural defects such as chevrons and edge dislocations in the $N_{TBF}$ phase of two individual compounds abbreviated JK103 (16), JK203 and a multicomponent mixture NTBF005. Details of the synthesis of JK203 and preparation of the mixture NTBF005 will be given elsewhere. All materials exhibit four nematic phases on cooling: a paraelectric nematic N, an antiferroelectric phase $N_x$, a ferroelectric nematic $N_F$ and a twist-bend ferroelectric nematic $N_{TBF}$. All three materials show a pronounced temperature dependence of the pitch $\mathcal{P}$ and the conical angle $\theta$ in the $N_{TBF}$ phase. Although the three materials exhibit qualitatively similar behavior, their quantitative differences facilitate a thorough characterization of many aspects of structural organization. For example, the pitch of JK203 is slightly larger than 1 µm, which helps in the optical characterization of periodic textures, while the room temperature mixture NTBF005 allows one to perform atomic force microscopy and characterization of chevron structures by rotation of cells.

The paper is organized as follows. Section II presents the phase sequences and the temperature dependencies of the material parameters such as pitch, conical angle, and polarization. This section also describes materials for surface alignment and methods of characterization. Section III.1 describes how a unidirectional surface polar axis **R** created by mechanical rubbing of the alignment layer of cell affects the orientation of polarization vector **P** in the $N_F$ and of the twist-bend axis **q** in the $N_{TBF}$: **P** in the $N_F$ is antiparallel to **R**, while **q** in the $N_{TBF}$ tilts away from $-\mathbf{R}$. The latter is attributed to surface reconstruction of the polarization field, driven by the avoidance of bound charge. The surface reconstruction produces optical activity along the cell normal in the $N_{TBF}$ phase, discussed also for tangentially degenerate cells in Section III.2, and is ultimately responsible for the formation of tilted and chevron $N_{TBF}$ structures as the surface and bulk periods of the structure differ from each other (Section III.3). The tilted and chevron structures are revealed in the experiments on light transmission through sandwich cells for different angles of incidence. Section III.4 describes edge dislocations that mediate temperature-induced variations of $\mathcal{P}$. Fitting the distortions of pseudolayers around the dislocation core with the nonlinear elastic model demonstrates that the elastic penetration length of the $N_{TBF}$ is on the order of $\mathcal{P}$. The dislocation core differs from its counterparts in paraelectric $N^*$ and ferroelectric $N_F^*$, as it does not split into semi-integer disclinations because of the global polarity condition $\mathbf{q} \neq -\mathbf{q}$. The absence of core splitting enables a high gliding mobility of dislocations. Mapping the optical retardance around the core demonstrates that the compression/dilation of pseudolayers changes the local conical tilt angle. Finally, Section IV summarizes the results.

## II.    MATERIALS AND METHODS

### II.1. Phase Sequences.

The phase sequences of the studied materials on cooling are presented below:

JK103, 4'-(difluoro(3,4,5-trifluorophenoxy)methyl)-2,3',5'-trifluoro-[1,1'-biphenyl]-4-yl 2,6-difluoro-4-(5-propyl-1,3-dioxan-2-yl)benzoate, **Figure 1**:

N → 149°C → $N_X$ → 147°C → $N_F$ → 103°C → $N_{TBF}$ → 92°C → $SmC_F$ ;

JK203, 4'-(difluoro(3,4,5-trifluorophenoxy)methyl)-2,3',5'-trifluoro-[1,1'-biphenyl]-4-yl 2',3,5,6'-tetrafluoro-4'-(5-propyl-1,3-dioxan-2-yl)-[1,1'-biphenyl]-4-carboxylate, **Figure 2**:

I → 327 °C → N → 240 °C → $N_X$ → 239 °C → $N_F$ → 161 °C → $N_{TBF}$ → 155 °C → $SmC_F$ → 148 °C → Crystal,

NTBF005, **Figure 3**: I → 176 °C → N → 133°C → $N_X$ → 131°C → $N_F$ → 30°C → $N_{TBF}$.

Here, I stands for the isotropic liquid phase, $N_X$ is the nematic phase with antiferroelectric block structure, also known as $SmZ_A$ (29) and $SmC_F$ is the ferroelectric smectic C. $T_c$ is the $N_F$ → $N_{TBF}$ transition temperature. The phase sequences are supported by differential scanning calorimetry (DSC) data and X-ray diffraction data, previously published for JK103 (16). The DSC scans for JK203 and NTBF005 are presented in **Figure S1** and **S2**, respectively. The X-ray diffraction data for JK203 and NTBF005 are presented in **Figure S3** and **S4**, respectively. It is important to stress that the X-ray diffraction patterns for the $N_{TBF}$ reveal the absence of long-range positional order and thus confirm the nematic nature of the phase, in which the pseudo-layering is caused by the periodic heliconical orientation of the molecules rather than by the density wave in their arrangements, Figure 1e. In contrast, the $SmC_F$ phase of JK103 and JK203 does show a long-range positional order, see Figure S3, and represents the new type of a ferroelectric smectic with the local polarization vector along the director rather than perpendicular to it.

Cooling of JK203 from the $N_F$ into the $N_{TBF}$ yields a "fingerprint" texture visible under a polarizing microscope, thanks to a relatively long pitch $\mathcal{P} = 1.2$ µm near the $N_F \to N_{TBF}$ phase transition. NTBF005 shows the $N_{TBF}$ phase at the room temperature, which facilitates exploration with an obliquely impingent light beam to demonstrate chevron defects and to visualize the periodic $N_{TBF}$ structure by atomic force microscopy. JK103 is used to study the alignment and edge dislocation of $N_{TBF}$ phase in cells with azimuthally degenerate anchoring.

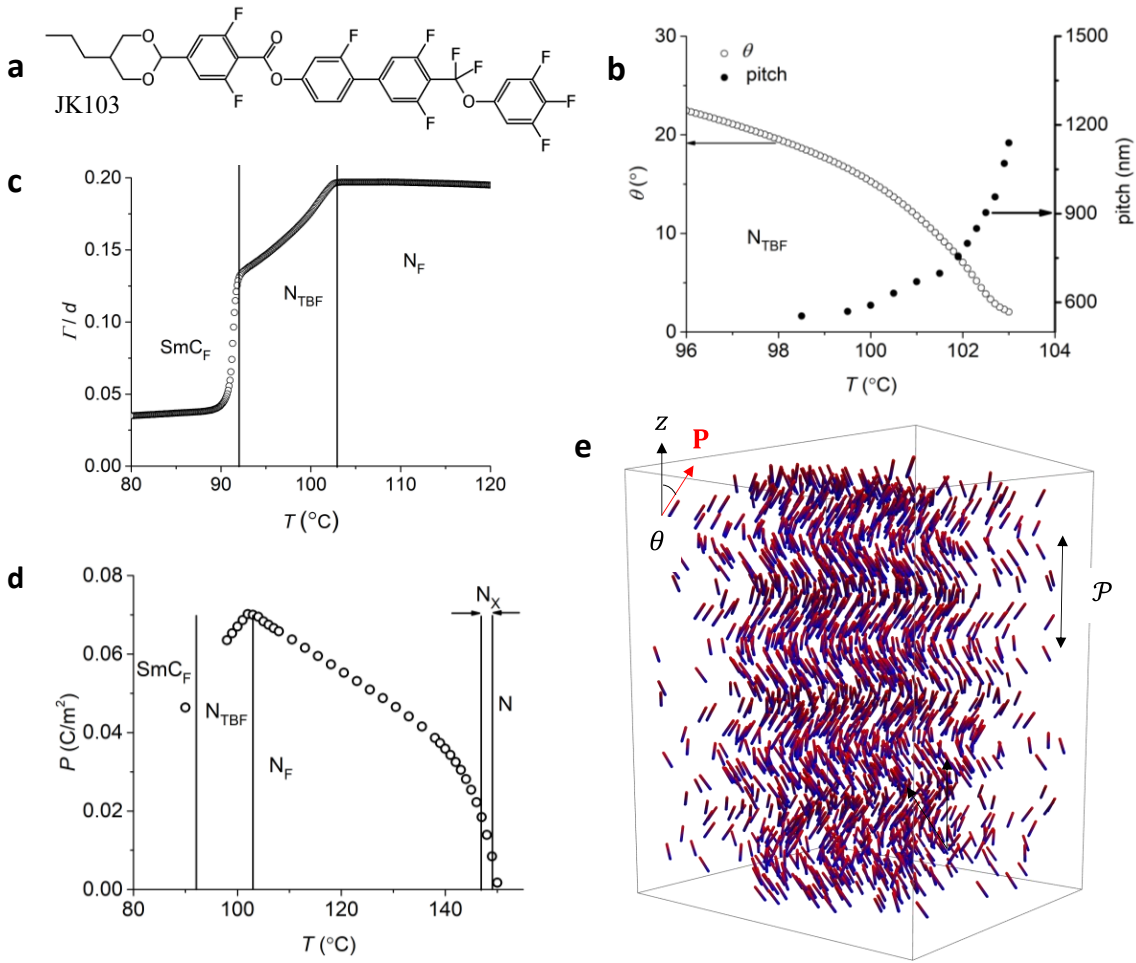

**Figure 1.** JK103: a) chemical structure, b) temperature ($T$) dependencies of pitch $\mathcal{P}$ and conical angle $\theta$ in the $N_{TBF}$ phase; c) temperature dependencies of optical retardance $\Gamma/d$ normalized by the cell thickness $d$ and d) of polarization $P$; e) Schematic structure of polar molecular organization in the $N_{TBF}$ illustrating a heliconical polarization field, Equation 1.

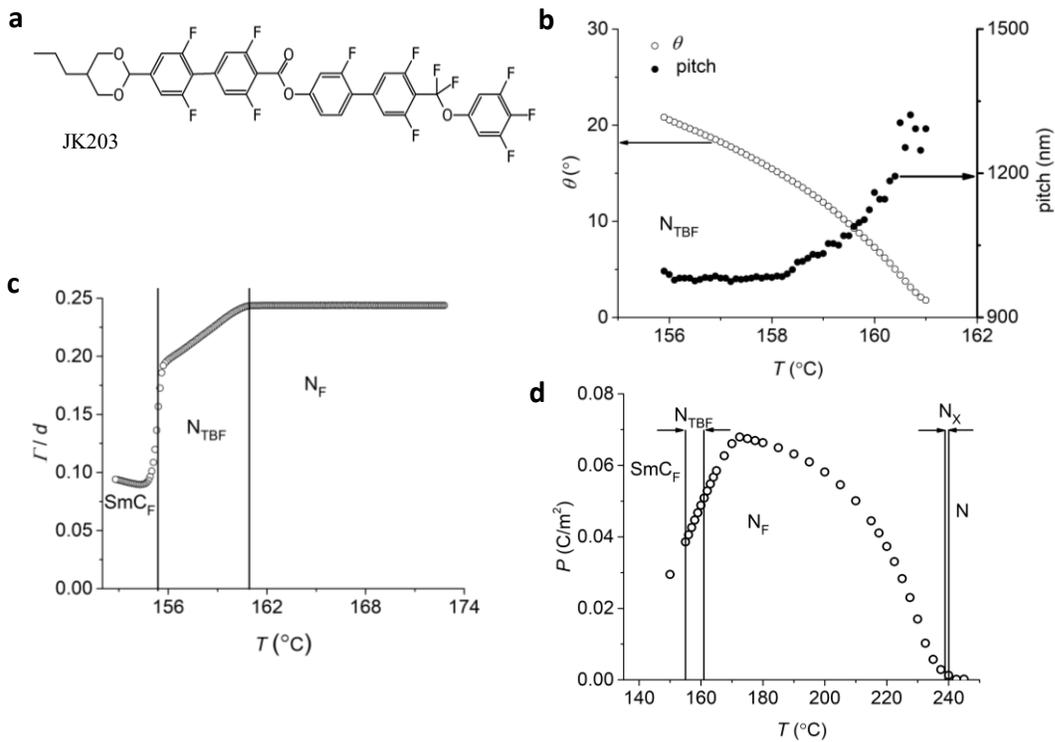

**Figure 2.** JK203: a) chemical structure; b) temperature ($T$) dependencies of pitch $\mathcal{P}$ and conical angle $\theta$ in the $N_{TBF}$ phase; c) temperature dependencies of optical retardance $\Gamma/d$ normalized by the cell thickness $d$ and d) of the polarization $P$.

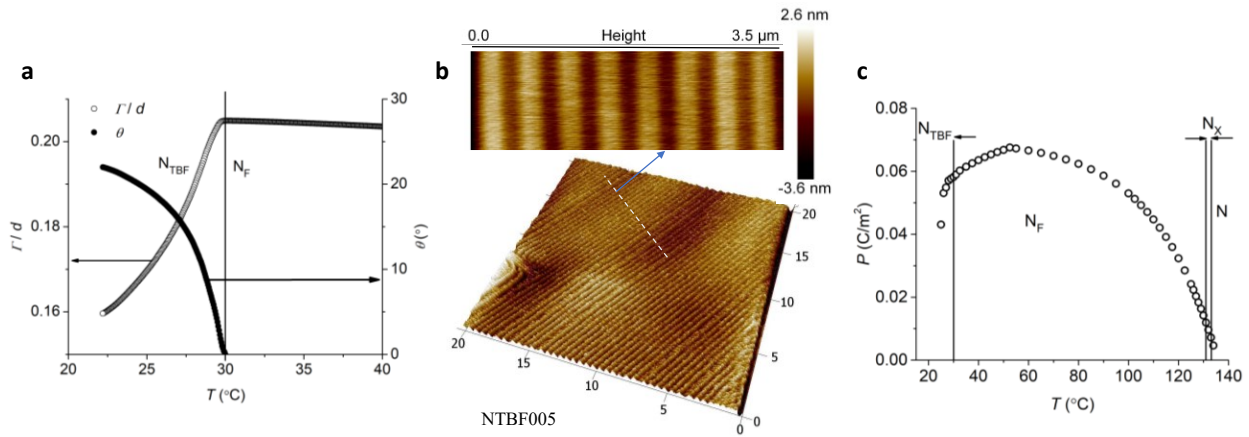

**Figure 3.** NTBF005: a) temperature ($T$) dependencies of optical retardance $\Gamma/d$ normalized by the cell thickness $d$ and of conical angle $\theta$ in the $N_{TBF}$; b) AFM texture of $N_{TBF}$ phase showing the surface period $\approx 0.5$ μm; c) polarization $P$ vs. temperature.

## II.2. Temperature dependencies of material properties.

To characterize the birefringence, we measure the optical retardance $\Gamma$ as a function of temperature in planar cells of a known thickness $d$ by the setup based on the Photoelastic Modulator (PEM-90, Hinds Instruments), for the green light ($\lambda = 532$ nm). The birefringence $\Delta n_{NF}$ of the $N_F$ and $\Delta n_{NTBF}$ of the $N_{TBF}$ is calculated as the ratio $\Gamma/d$, Figure 1c, 2c, 3a. The conical angle $\theta$ is estimated by using the expression (15) $\Delta n_{NTBF} \approx \Delta n_{NF}\left(1 - \frac{3}{2}\theta^2\right)$, Figure 1b, 2b, 3a. Figure 1e shows the schematic structure of $N_{TBF}$ and defines $\theta$ and $\mathcal{P}$.

The pitch $\mathcal{P}$ is measured by light diffraction experiments described previously (16), Figure. 1b, 2b; in some cases, the helicoidal structure and its pitch can be directly visualized by atomic force microscopy, Figure 3b.

We also measure the temperature dependencies of polarization $P$ by applying a weak alternating current (ac) electric field, of a triangular waveform with a peak-to-peak amplitude $2.4 \times 10^5$ V/m, across a 1 mm gap between two electrodes in the plane of the cell, **Figure S5** and **S6**. The polarization reaches a maximum value $P \approx 0.07$ C/m$^2$ in the $N_F$ phase and decreases when the materials are cooled down into the $N_{TBF}$ phase, Figure 1d, 2d, 3c, since **P** forms a heliconical structure with a finite cone angle $\theta$, Figure 1e. As described previously (16), application of a strong electric field, $5 \times 10^5$ V/m, along the normal to a $N_{TBF}$ cell increases the polarization by full realignment of **P** along the field, $\theta \to 0$. For comparison, we measured the temperature dependency of $P$ for the well-studied $N_F$ material DIO, **Figure S7**, which equals 0.054 C/m$^2$ at 50 °C; the results are close to those reported by Nishikawa et al. (2) and Chen et al. (29).

## II.3. Surface alignment agents

a. <u>Unidirectionally rubbed polyimide PI2555 coatings</u>

A PI2555 alignment layer of a thickness of 50 nm is spin-coated onto the glass substrates following Ref. (30). The PI2555 layer is unidirectionally rubbed using a Rayon YA-19-R cloth (Yoshikawa Chemical Company, Ltd, Japan) of a thickness of 1.8 mm and filament density $280/\text{mm}^2$. An aluminum brick of a length of 25.5 cm, width of 10.4 cm, height of 1.8 cm, and weight of 1.3 kg, covered with the cloth, imposes a pressure of 490 Pa at a substrate and is moved unidirectionally ten times with a speed of 5 cm/s over the substrate; the rubbing length is about 1 m (30). The rubbed PI2555 typically aligns the director of a paraelectric N parallel to the rubbing direction **R** with a small pretilt $\approx 3° \pm 1°$ (31). In the Cartesian coordinates, the plane $xz$ is parallel to the bounding plates, the $z$-axis is antiparallel to **R**, and the $y$-axis is normal to the cell. In the $N_F$ phase of all three materials, the rubbing direction $\mathbf{R} = (0,0,-1)$ aligns the spontaneous polarization antiparallel to itself, $\mathbf{P} = P(0,0,1)$, **Figure S8**, similarly to DIO formed by fluorinated molecules (30, 32). This behavior is opposite to RM734 molecules with $NO_2$ end groups, in which case $\mathbf{P} = P(0,0,-1)$ is parallel to the rubbing direction, $\mathbf{P} \downarrow\downarrow \mathbf{R}$ (33).

b. <u>Unrubbed polystyrene coatings</u>

Glass substrates are cleaned ultrasonically in distilled water and isopropyl alcohol, dried at 95 °C, cooled down to room temperature, and blown with nitrogen. Spin coating is performed using a 1 % solution of polystyrene in chloroform for 30 s at 4000 rpm. After the spin coating, the sample is baked at 45°C for 1 h. Unrubbed polystyrene coatings typically set in-plane degenerate anchoring for the N (34) and $N_F$ (35).

c. <u>Cell assembly.</u>

Two PI2555-covered glass substrates form planar cells. The rubbing directions $\mathbf{R} = (0, 0, -1)$ on the plates are parallel to each other. The cell gap thickness $d$ is set by glass spherical spacers of a diameter in the range 2-10 µm; $d$ is measured at five different locations within a cell by an interferometric technique using UV/VIS spectrometer Lambda 18 (Perkin Elmer). The standard variation of $d$ is 0.1 µm or less. The liquid crystal is filled into cells in the N phase by the capillary flow along **R**. A Linkam hot stage controls the temperature with an accuracy of $\pm 0.1°C$. Cells with azimuthally degenerate anchoring are prepared from glass substrates with polystyrene coatings as described previously (36).

**II.4. Polarizing optical microscopy.**

A polarizing optical microscope Olympus BX51 equipped with a Basler video camera is used for textural observations. The optical retardance vs. temperature was measured with a setup based on the Photoelastic Modulator (PEM-90, Hinds Instruments), for the green light ($\lambda = 532$ nm). Surface mapping of retardance was performed with the PolScope MicroImager (Hinds Instruments). Analysis of optical activity was performed in monochromatic light with a green interferometric filter (center wavelength $\lambda = 532$ nm, bandwidth 1 nm).

## II.5. Atomic force microscopy.

AFM measurements were performed using a Bruker Dimension Icon Microscope working in modified tapping ScanAsyst mode and cantilevers with 0.4 N/m force constant were applied.

## II.6. Statistical Analysis

Optical retardance measurements: The optical retardance $\Gamma$ is measured using Photoelastic Modulator (PEM-90, Hinds Instruments) with the accuracy of $\pm 1.0$ nm. The measured retardance $\Gamma=(975.0 \pm 1.0)$ nm of the cell of gap thickness $5.0\ (\pm 0.1)$ µm filled with JK103, Figure 1c, yields the optical birefringence of $0.195\pm 0.004$. The contribution of the instrumental errors including the measurements of the gap thickness and retardance to the birefringence uncertainty is less than $\pm 2\%$.

Transmitted light intensity measurement: The intensity $T$ of transmitted light in the optical microscopy textures is determined by using ImageJ software. The error bar in transmission $T$ is less than 1%. The tilted and chevron structures of $N_{TBF}$ phase are studied using a setup with a photodetector. The light intensity transmitted through a pair of crossed polarizers without any sample represents dark noise of the setup. The reported values of light intensity in textures are higher than the dark noise by a factor 100 or larger. The instrumental error in the measurements of $T$ is less than 1 %.

Polarization density measurment: The measurement of polarization $P$ involves the measurement of the electric current through a cell and the cross-sectional area of cells. The electric current through a 20 k$\Omega$ resistor is determined by using an oscilloscope Tektronix TDS 2014 (sampling rate 1GSa/s). The voltage across the resistor is measured with the accuracy of $10^{-3}$ V, which yields the accuracy of $5 \times 10^{-8}$ A for the current. The polarization peak current is higher than $5 \times 10^{-6}$ A. The error in the measurement of current is less than 1%. The integral of current over time is calculated using numerical approximation of area under the curve in Origin software; the accuracy is better than 0.1%. Calculations of the cross-section area $A$ of cells involves the measurement of the cell gap thickness and length of the electrodes. The error in $A$ measurements is less than 0.5%. The error in the polarization density measurment including all parameters explained above is less than 2%.

## III. RESULTS

### III.1. Surface alignment in planar cells with parallel assembly.

a. <u>$N_F$ alignment.</u>

Planar cells with two surfaces covered with unidirectionally rubbed PI2555 in parallel assembly show a uniform texture of the N phase with the director along the rubbing direction **R**. The intermediate $N_X$ phase in all three materials exhibits polydomain textures with zig-zag domain walls, similar to the textures of the antiferroelectric $SmZ_A$ phase of DIO (29). We do not observe large-scale optically resolved stripes reported by Rupnik et al. (37) for the intermediate phase of RM734 mixed with an ionic fluid, nor the square lattices reported by Ma et al. (38) for an intermediate phase in the presence of ionic polymers. An apparent reason is the strong polarization of the explored materials, which suppresses the flexoelectric effects responsible for the splay tendency (29). Slow cooling of the antiferroelectric $N_X$ to the ferroelectric $N_F$ phase in planar cells of all three materials results in a uniform alignment of polarization that is antiparallel to the rubbing direction, **P**↑↓ **R**, as established by switching in the in-plane electric field, Figure S8.

b. <u>$N_{TBF}$ alignment.</u>

Cooling down a planar cell into the $N_{TBF}$ yields a "fingerprint" texture in which the twist-bend axis **q** makes an angle different from 180° with **R**, **Figure 4a,** as verified by applying an in-plane electric field. The textures often reveal edge dislocations, which accommodate the cooling-induced reduction of $\mathcal{P}$ by introducing new pseudolayers, Figure 4a. The term pseudolayers stresses the absence of density modulation in the $N_{TBF}$. The Burgers vector **b** equals $\mathcal{P}$ in magnitude and is collinear with **q**. The edge dislocations climb along the pseudolayers, thus adjusting a new equilibrium $\mathcal{P}$.

Why **q** is not collinear with **R**, Figure 4a? An unperturbed $N_{TBF}$ with **q** antiparallel to **R**, **q**↑↓ **R**, is difficult to accommodate since it creates surface charges and depolarization fields in the surface regions where **P** tilts away from the $xz$ plane, Figure 4b. To avoid the charges, **P** should reorganize by reducing the $y$ component, $P_y \to 0$, at the interfaces, Figure 4c. As one moves into the bulk, a merger with the ideal heliconical **P** requires splay and twist, Figure 4c. The concrete mode of reconstruction depends on the balance of electrostatic, anchoring, and elastic energies;

deciphering this balance is a complex problem with solutions available for the $N_F$ (39, 40) but not for the $N_{TBF}$. However, the very existence of twist along the $y$-axis normal to the cell is revealed in an optical activity experiment, **Figure 5**.

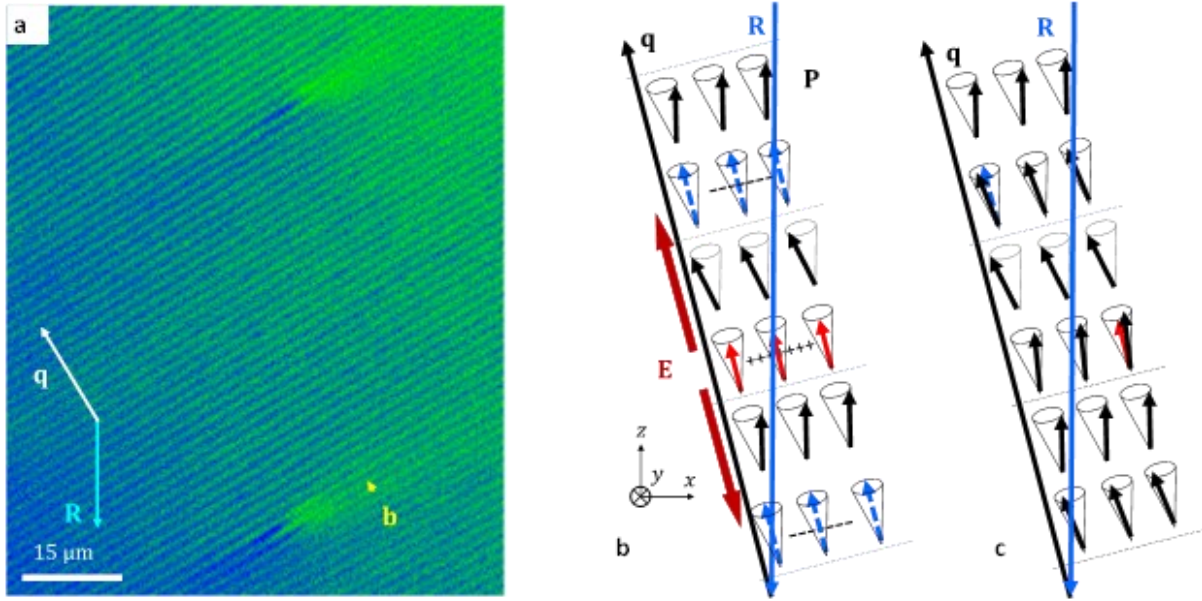

**Figure 4.** Periodic fingerprint texture of JK203 $N_{TBF}$ below the transition point from the $N_F$, $T = T_C - 0.1\,°C$: a) The heliconical axis **q** in the optical microscope texture is not collinear with the rubbing direction **R**. There are two edge dislocations in the texture, of the Burgers vector **b** along **q** and of a magnitude $\mathcal{P}$. b) Scheme of an unperturbed heliconical $N_{TBF}$ superimposed onto the $xz$ plane with **R** along one of two tangential orientations of the $N_{TBF}$ molecules, $P_y = 0$. Tilts of **P** towards the positive end of the $y$-axis (blue arrows) and the negative end (red arrows) deposit negative and positive surface charges, respectively, thus creating a depolarization electric field **E**. Black arrows show **P** tangential to the $xz$ plane. c) Schematic surface restructuring of the $N_{TBF}$ in which the surface **P** is everywhere in the $xz$ plane; as one moves into the bulk along the $y$-axis, the surface polarization merges with the ideal heliconical bulk structure through splay and twist, as illustrated by co-localized red and blue bulk arrows with the black surface arrows of **P**.

The optical activity is characterized in areas free of dislocations, Figure 5a. The crossed polarizer P and analyzer A are rotated while remaining orthogonal, Figure 5b. The transmitted light intensity is measured as a function of the angle $\Phi$ between the polarizer and **R**. In the N and $N_F$ phases, the minimum transmission is at $\Phi_m = 0$ since the optic axis (i.e., the director and **P**) is along **R**. In the $N_{TBF}$, the location $\Phi_m$ of the transmission minimum increases from 0 to $\approx 15°$ when the temperature lowers from $T_C$ to $T_C - 2.5°C$, indicating that the optic axis tilts away from **R**, Figure 5b. Light transmittance measured as a function of the angle $\gamma$ between the polarizer and analyzer, when the polarizer direction is fixed at $\Phi_m$, presents an independent proof of optical activity, Figure 5c,d,e,f. At $T_C$, the transmission follows the dependence $T \propto \cos^2 \gamma$, signaling that the optical axis is parallel to **R**. At $T_C - 0.5\ °C$ and below, there is a noticeable light leakage at $\gamma = 90°$ and a shift of the transmission minimum from the $\gamma = 90°$ location, Figure 5f, which demonstrates optical activity and twist of **P** along the $y$-axis.

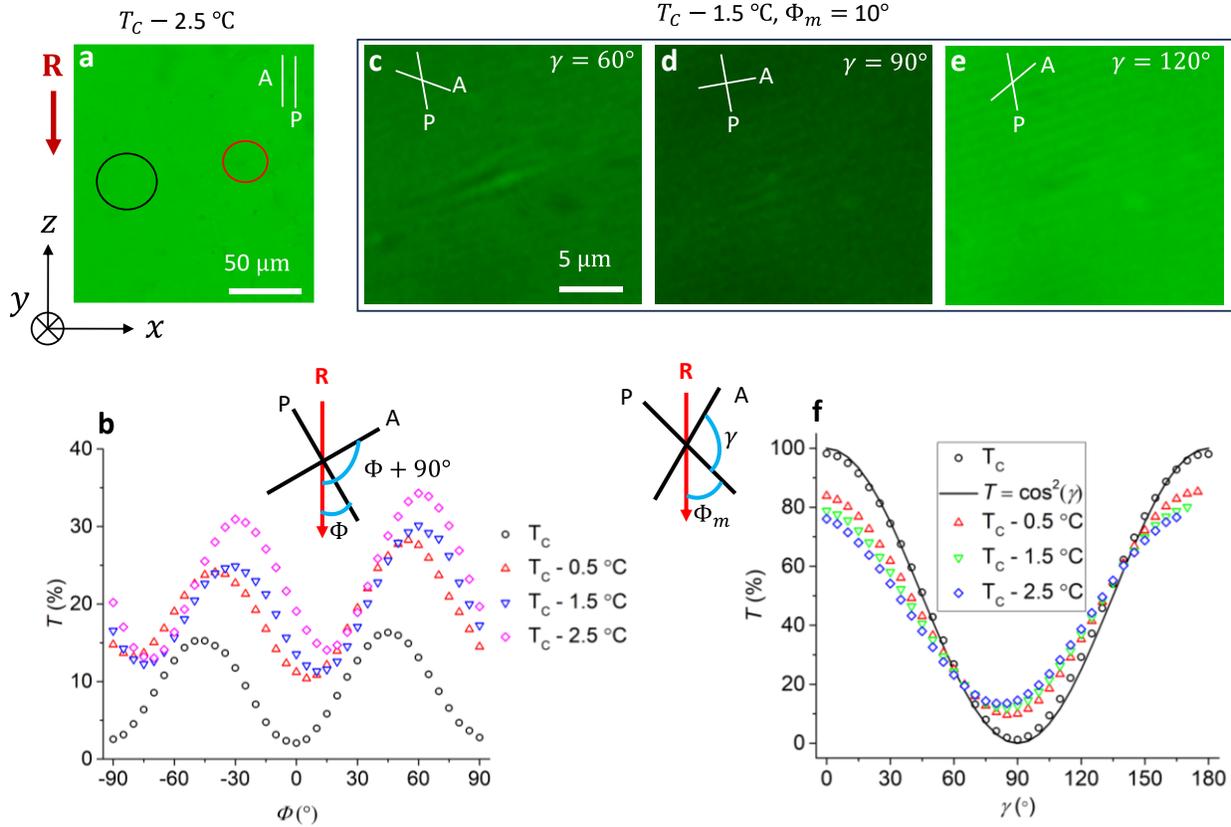

**Figure 5.** Surface reconstruction and optical activity in planar JK203 $N_{TBF}$ cell: a) Polarizing microscope texture for parallel polarizer P and analyzer A in the $N_{TBF}$, observed with a green interferometric filter. Red and black circles enclose an edge dislocation and a dislocation-free area, respectively. The dislocation-free area is used for optical characterization. b) Light intensity transmitted through the cell and crossed polarizers vs. the angle $\Phi$ between the polarizer and **R**. Note the shift of the minimum in transmittance at the temperatures below $T_C$. c), d), e) Polarizing microscope textures for the polarizer and analyzer making an angle $\gamma = 60°$, $\gamma = 90°$, and $\gamma = 120°$, respectively, $\Phi = \Phi_m = 10°$. The difference in transmittance for $\gamma = 60°$ and $\gamma = 120°$ suggests optical activity. f) Transmitted light intensity $T$ as a function of $\gamma$; the polarizer is fixed at $\Phi = \Phi_m$. At temperatures $T_C - 0.5\ °C$, optical activity is evidenced by light leakage at $\gamma = 90°$ and a shift of transmission minimum from $\gamma = 90°$. Cell thickness $d = 6.6\ \mu m$.

## III.2. Surface alignment in cells with azimuthally degenerate anchoring.

Optical activity along the normal to the $N_{TBF}$ cell with the twist-bend **q** axis in the plane of the cell is a general feature observed not only in rubbed PI2555 planar cells but also in cells with unrubbed polystyrene coatings that produce in-plane degenerate anchoring for the N (34) and $N_F$ (36). In these cells, the $N_F$ forms circular vortices of polarization, to avoid space charge (35, 36, 41). The vortices are bounded by hyperbolic and parabolic domain walls (35, 36, 41), **Figure 6a**. When the sample cools down to the $N_{TBF}$, it shows optical activity, as evidenced by the complementary contrast of images taken with left- and right-handed decrossings of the polarizer and analyzer, Figure 6b-d.

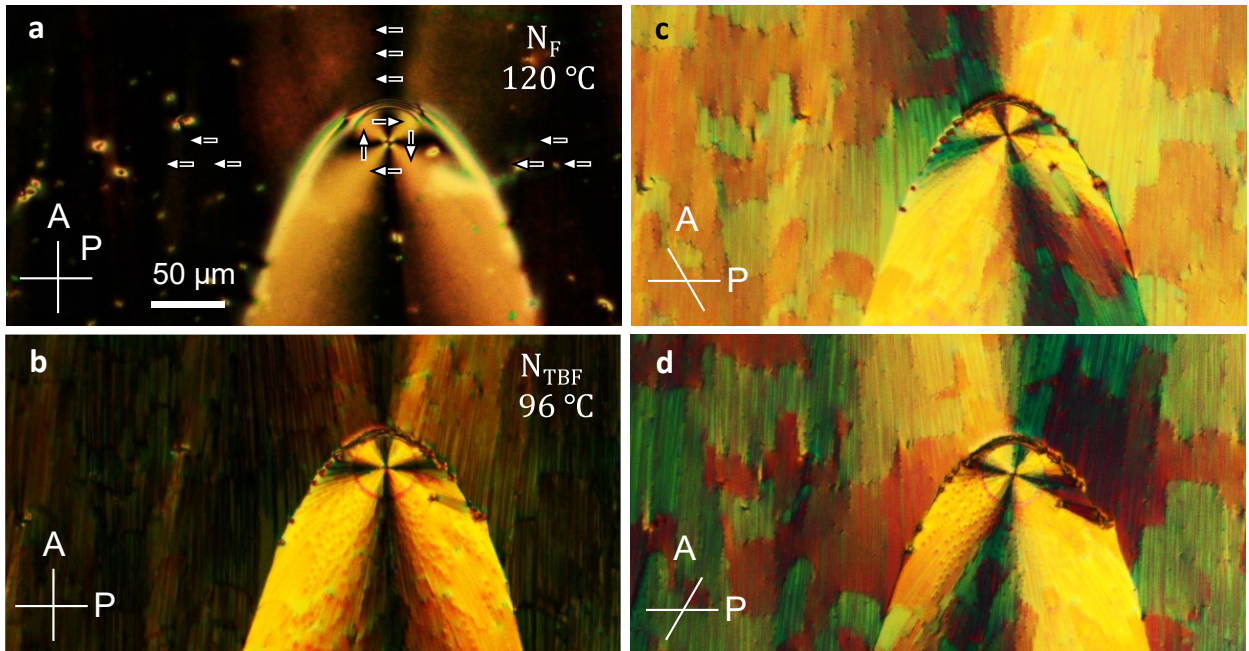

**Figure 6.** Optical activity of the JK103 $N_{TBF}$ cell with polystyrene coatings that produce azimuthally degenerate anchoring: a) A circular bend domain separated from a nearly uniform domain by a parabolic domain wall in the $N_F$ phase; arrows show the polarization field; b) The same region in the $N_{TBF}$ phase; c), d) optical activity of the $N_{TBF}$ texture is evidenced by complementary contrast of textures observed with left- and right-handed decrossings of the polarizer P and analyzer A, respectively. Cell thickness $d = (6.1 \pm 0.1)$ μm.

The optical activity in $N_{TBF}$ observed in both unidirectionally rubbed, Figure 5, and azimuthally-degenerate cells, Figure 6, is caused by the subsurface twists, which transform the linear polarization of light entering the sample into an elliptical polarization of the transmitted light. These twists are associated with the surface restructuring, which reduces the $y$-component of polarization and thus rotates **P** over the cone. The azimuthal direction of **P** at a cone sitting at the interface is thus different from that at a cone in the bulk. In the N and $N_F$, there is no heliconical structure, the optic axis is along **R** and there is no optical activity. The $N_{TBF}$ tilt angle $\theta$ increases as the temperature is lowered (16), which leads to stronger electrostatic frustration at the surface and thus stronger splay-twist deformations in surface reconstruction; the chevrons and tilted structures grow stronger when the temperature drops below $T_C - 0.2\ °C$.

### III.3. Uniformly tilted/chevron structures of N<sub>TBF</sub> phase.

Layered liquid crystals such as smectic C (42-44) and smectic A (45, 46) in flat planar cells often adopt the so-called chevron structure, in which the layers tilt away from the normal to the cell, bending in the middle. In thin cells, the layers show a unidirectional tilt. The structures result from the mismatch of the surface and bulk periods. The mismatch might be produced by the temperature dependence of the period or by the surface-imposed microscopic interactions. Both these factors are present in the $N_{TBF}$, as $\mathcal{P}$ shows a strong temperature dependence, Figure 1-3, and the surface reconstruction distorts the ideal heliconical structure, Figure 4. Measurements of light transmission through a planar sample that is rotated around the $x$-axis, **Figure 7a**, demonstrate that depending on the cell thickness, the $N_{TBF}$ samples form uniformly tilted, Figure 7b, and chevron, Figure 7c, structures when cooled from the $N_F$ phase, **Figure 8**. The light transmittance through the tilted cells is an optical analog of the X-ray scattering experiments used to discover chevrons in smectics with nanoscale periodicity of material density (42-47). Since the $N_{TBF}$ lacks long-range positional order, Figure S3 and S4, optical methods are much better suited for the demonstration of chevron and titled $N_{TBF}$ structures as compared to the X-ray diffraction.

The transmitted intensity of the linearly polarized He-Ne laser beam, 632.8 nm wavelength, is measured as a function of incident angle $\alpha$; $\alpha = 90°$ represents normal incidence, Figure 7a. Tilts of the cell in opposite directions ($\alpha < 90°$ and $\alpha > 90°$) produce asymmetric response with a maximum that is shifted from $\alpha = 90°$, indicating that the pseudolayers are tilted away from the $y$-axis. Within the cell, the maxima of transmission can be located at both $\alpha < 90°$ and $\alpha > 90°$, Figure 8d,e, which means that the sample is split into domains in which the vector **q** tilts up and down along the $y$-axis, Figure 7b,c. A single maximum is observed in cells thinner than some critical thickness $d_{ch} = (7-9)$ µm; it corresponds to uniformly tilted pseudolayers, Figure 7b. Thicker cells, $d > d_{ch}$, show two maxima that correspond to a chevron, since both directions of the tilt are present along the $y$-axis, Figure 7c, 8f. One observes coexisting chevrons with bend regions oriented in opposite directions, Figure 7c; these chevrons are separated by zig-zag domain walls, Figure 8c. Similar zig-zag walls are observed in smectic chevron textures (47). Note that qualitatively the same transmittance dependence on $\alpha$, with one shifted maximum in thin cells and two maxima in thick cells, is observed for other types of light polarization, including circular polarization, which indicates that optical activity does not obscure the pseudolayers' tilts.

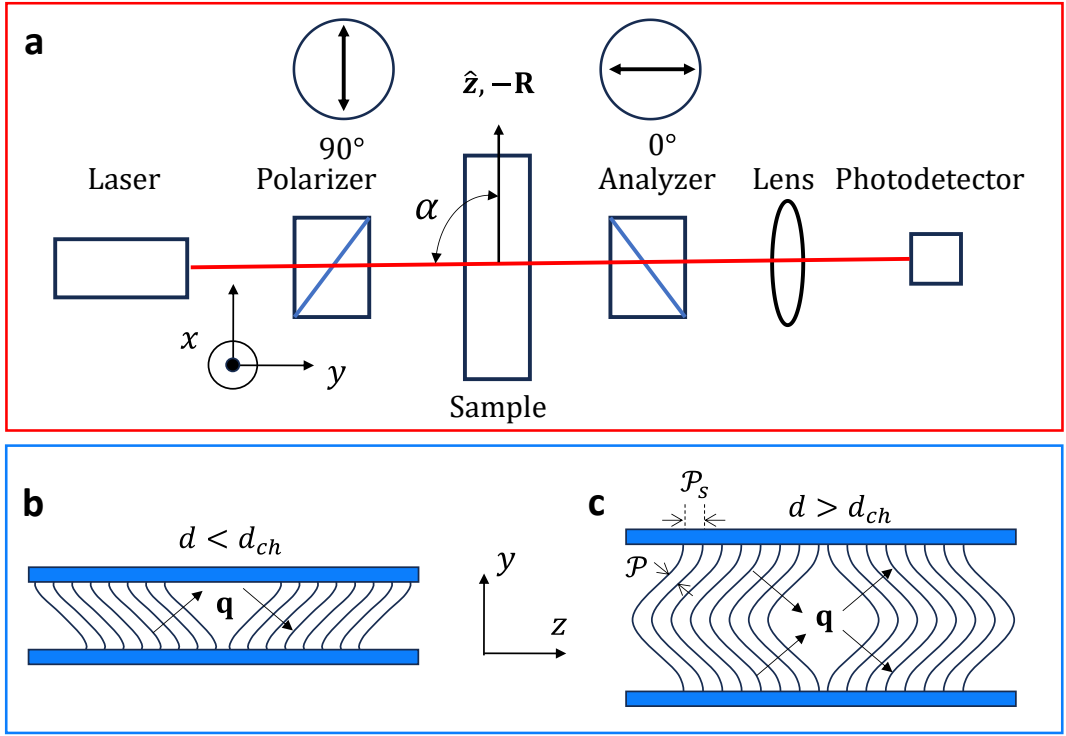

**Figure 7**. Tilted and chevron arrangements of $N_{TBF}$ pseudolayers: a) Experimental setup. The cell is tilted around the $x$-axis by an angle $\alpha$; $\alpha = 90°$ represents the normal incidence of the light; b) scheme of tilted pseudolayers in thin cells; c) chevron structure in thick cells.

The tilted and chevron structures can be explained by the Limat-Prost model, which was developed for one-dimensionally periodic structures (48). The model considers the stress $\gamma_s$ caused by a difference of bulk and surface periodicities. In our notations, $\gamma_s = (\mathcal{P}_s - \mathcal{P})/\mathcal{P}$, where $\mathcal{P}_s$ is the surface period, Figure 7c. When $\gamma_s$ exceeds a critical value $\gamma_{st} = \pi^2\lambda^2/d^2$, a tilted structure forms; when $\gamma_s$ exceeds a higher threshold

$$\gamma_{sc} = 4\pi^2\lambda^2/d^2, \qquad (2)$$

a chevron forms. Here $\lambda = \sqrt{K/B}$ is the elastic penetration length, $K$ is the curvature Frank constant, and $B$ is the compressibility modulus. The critical stress can be estimated by assuming that $\mathcal{P}_s$ equals the pitch $\mathcal{P}$ at the transition point $T_c$. Then in the deep $N_{TBF}$, as follows from Figure 1b, $\gamma_s$ can be as high as 0.5. As will be shown later, $\lambda \approx \mathcal{P}$. For $\lambda = 1$ μm, the critical stress for the tilted structures is achieved at $d \geq 4$ μm, while the chevron is expected to form when $d \geq 18$ μm. These numbers are in reasonable agreement with the experiment in Figure 8. If the surface interactions make $\mathcal{P}_s$ larger than $\mathcal{P}$ at $T_c$, then the agreement will further improve.

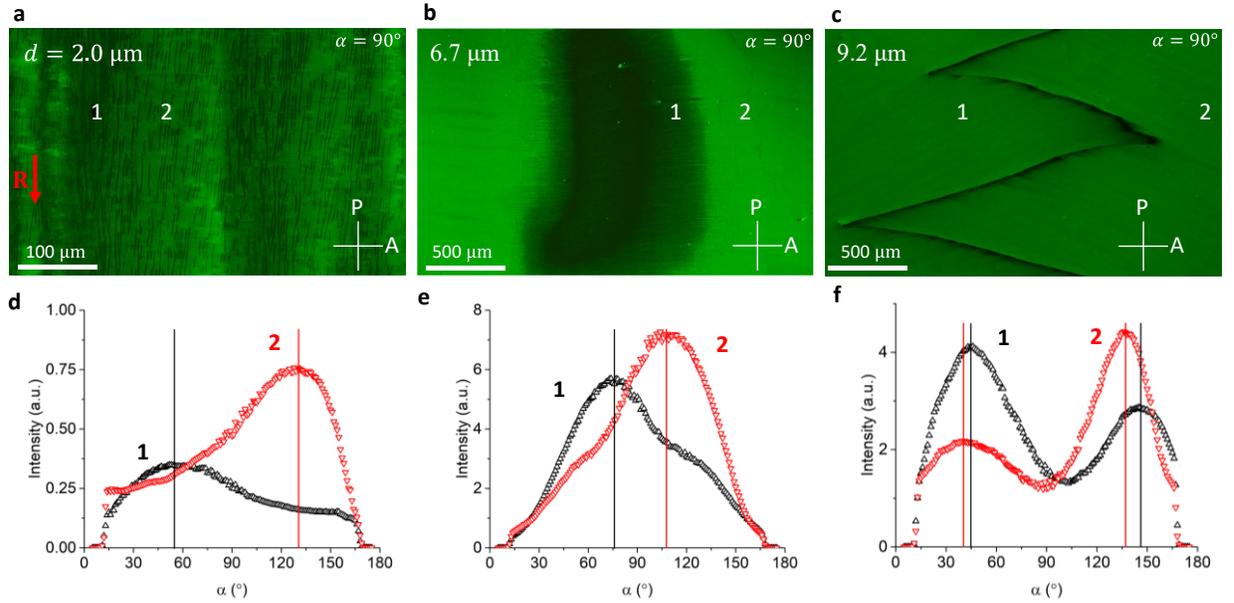

**Figure 8.** Light transmission through planar cells with uniformly tilted and chevron structures of NTBF005 $N_{TBF}$ pseudolayers: a), b), c), Polarizing microscope of the $N_{TBF}$ texture in cells of thickness $d = 2.0$ μm, 6.7 μm, and 9.2 μm, respectively; 22 °C or $T_C - 8$ °C; monochromatic 532 nm probing light; note the secondary stripes in thin cell in a); d), e), f), Transmitted light intensity vs. the tilt angle $\alpha$ at the locations 1 and 2 of a), b), c). A single peak in d), e) reveals a tilted structure; double peaks of different amplitudes in f) correspond to an asymmetric chevron.

An interesting feature of thin cells, $d = 2.0$ μm, Figure 8a, is the presence of "secondary" stripes which appear to be nearly perpendicular to the pseudolayers. These might be caused by in-plane undulations. In thin cells, the surface anchoring might be strong enough to keep the period close to $\mathcal{P}_s$ across the entire cell. If the equilibrium pitch decreases, the stress can be relieved by tilting **q** left and right in the $xz$ plane of the cell, producing a chevron that is orthogonal to the conventional chevron in Figure 7c. Such an orthogonal chevron forms in chiral smectic C (49).

### III.4. Edge dislocations

Changing the temperature of the sample changes $\mathcal{P}$, Figure 1b and Figure 2b. A decrease or increase of $\mathcal{P}$ is accommodated by adding new or removing existing pseudolayers, respectively. The process is enabled by edge dislocations, Figure 4a and **Figure 9-12**. Below, we describe these dislocations using the elastic model developed for a nonpolar smectic A. Such an approach has been used to describe the elasticity of $N^*$ (24, 50, 51), $N_F^*$ (28), and $N_{TB}$ (52-55). In the case of $N_{TBF}$, it is justified when the scale of deformations is much larger than $\mathcal{P}$, and when there is no electric field acting on the sample.

a. <u>Edge dislocations and penetration length.</u>

An elementary edge dislocation of a Burgers vector $b = |\mathbf{b}| = \mathcal{P}$, **b** ∥ **q**, represents the line end of a semi-infinite $N_{TBF}$ pseudolayer sandwiched between intact pseudolayers, Figure 4a and Figure 9. In what follows, we neglect the effect of surface reconstruction and tilts of pseudolayers discussed above and assume that the pseudolayers are along the $y$-axis. Within this assumption, the normal **q** to the pseudolayers in a dislocation-free sample is along the $z$-axis. A narrow temperature range between $T_c$ and $T_C - 0.2\ °C$ at which the tilts and chevrons are not yet formed, satisfies this assumption best.

The elastic energy functional for deformations developing over scales much larger than $\mathcal{P}$, depends on a single scalar variable, the displacement $u(x, z)$ of pseudolayers along the $z$-axis (56). The free energy density contains terms for weak bends and compressions/dilations:

$$f = \frac{1}{2}K\left(\frac{\partial^2 u}{\partial x^2}\right)^2 + \frac{1}{2}B\left[\frac{\partial u}{\partial z} - \frac{1}{2}\left(\frac{\partial u}{\partial x}\right)^2\right]^2, \quad (3)$$

The correction $\left[-\frac{1}{2}\left(\frac{\partial u}{\partial x}\right)^2\right]$ makes the compressibility term invariant with respect to uniform tilts of pseudolayers. With the elastic energy density in **Equation 3**, an analytical solution for the displacements $u(x,z)$ around an edge dislocation with the core at $(x,z) = (0,0)$ and $\mathbf{b} = (0,0,b)$ reads (57)

$$u(x,z) = 2\lambda \ln\left[1 + \frac{1}{2}\left(e^{\frac{b}{4\lambda}} - 1\right)\left(1 + \text{Erf}[\frac{x}{2\sqrt{z\lambda}}]\right)\right], \quad (4)$$

where Erf[...] is the error function defined as $\text{Erf}\left[\frac{x}{2\sqrt{z\lambda}}\right] = \frac{2}{\sqrt{\pi}} \int_0^{\frac{x}{2\sqrt{z\lambda}}} exp(-t^2) dt$. In the $N_{TBF}$, $\lambda$ is not known. To estimate it, we fit the pseudolayers around the dislocation core with **Equation 4** for different $b/\lambda$, as described in Ref. (58). As shown in Figure 9b, $b/\lambda = 1$ yields a much better fit as compared to $b/\lambda = 10$ and $b/\lambda = 0.1$. Therefore, in the explored JK203 $N_{TBF}$ phase, an order of magnitude estimate is $\lambda \approx b = \mathcal{P}$.

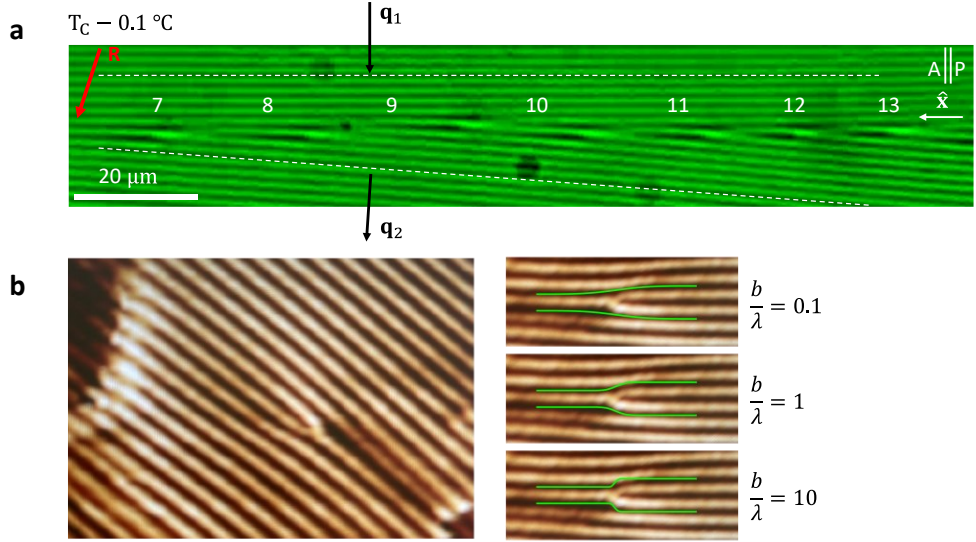

**Figure 9.** Dislocations in $N_{TBF}$ JK203: a) an optical microscope texture of a grain boundary with pseudolayers misaligned by a small angle $\beta$ in a planar cell. The numbers indicate the number of pseudolayers on the left- and right-hand sides of the dislocations within the region bounded by dashed lines. The polarizer P and analyzer A are parallel to each other. b) Atomic force microscopy texture of an isolated edge dislocation and fitting of the pseudolayers with Equation 4.

The result $\lambda \approx \mathcal{P} \approx 1$ μm allows one to estimate the compressibility modulus $B = K/\lambda^2 \approx 10$ N/m² if the Frank modulus is of a typical value $K = 10$ pN. This estimate can be recast (53, 54) as $B = q^2 K \theta^2 \approx 10$ N/m². The value $B \approx 10$ N/m² is dramatically smaller than the compressibility modulus $B \sim (10^6 - 10^7)$ N/m² in smectics A (59, 60) and C (61), $B \sim (10^3 - 10^6)$ N/m² in the $N_{TB}$ phase (62-64), and $B = 4 \times 10^4$ N/m² in the SmZ$_A$ phase (65), which should not be surprising since the $N_{TBF}$ period is $\sim 10^3$ times larger than the period of smectic phases, 100 times larger than the $N_{TB}$ period, and 50 times larger than the $SmZ_A$ period (29).

  b. Grain boundaries formed by edge dislocations.

An edge dislocation creates dilations/compressions that are partially relaxed by a wedge-like tilt of pseudolayers. This wedge attracts more dislocations forming a low-angle grain boundary, which separates two grains misaligned by an angle $\beta$ (66), Figure 9a. The latter can be calculated by measuring the separation $s$ of the dislocations. With $s \approx 19$ μm and $\mathcal{P} \approx 1.6$ μm for the marked dislocations in Figure 9a, one finds $\beta = \arctan(\mathcal{P}/s) \approx 0.08$ rad, or 4.8°.

The tendency of dislocations to assemble into grain boundaries is very pronounced in the cells with degenerate azimuthal anchoring, Figure 10. When the circular vortex of polarization in the $N_F$ is cooled down into the $N_{TBF}$, the formed pseudolayers need to run radially from the vortex center. If the pseudolayers keep their period $\mathcal{P}$ close to the equilibrium value, such a radial configuration is possible only when the edge dislocations introduce new pseudolayers as one moves away from the vortex core, Figure 10a. The dislocations assemble into grain boundaries that run along the radial directions, to reduce the overall elastic energy.

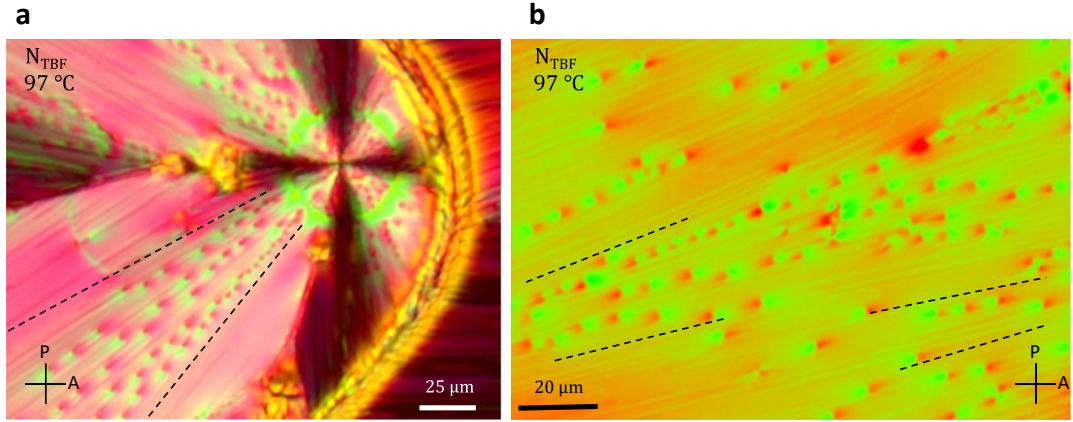

**Figure 10.** Polarizing optical microscopy texture of edge dislocations in $N_{TBF}$ JK103 cell with azimuthally degenerate anchoring: a) in a circular domain and b) in a nearly uniform domain. Dashed lines outline the splay of pseudolayers. Note a deterministic relationship between the direction of splay and the polarity of birefringence change at the dislocation cores: red color is always towards the compressed side. Cell thickness $d = 6.0$ μm.

The low-angle grain boundary can be modeled as if the dislocations were created in a wedge, with the left side being of height $h_N = Np$ and the right of height $Np + b$; $N$ is the number of pseudolayers. To find the equilibrium position of the dislocation cores and the separation between them, we align the $x$-axis along the pseudolayers of one domain; the neighboring domains are misaligned by an angle $\beta$.

The equilibrium position $x_d$ of the dislocation core can be found by calculating the compression energy (67). The compressibility energy density written in linear approximation as $f = \frac{1}{2}B\left(\frac{\partial u}{\partial z}\right)^2$, is integrated over the wedge area of a length $b/\tan\beta$, a height $h_N$ on the left side and $h_N + b$ on the right side:

$$E(x_d) = \frac{B}{2}\left[\int_{h_N/\tan\beta}^{x_d} dx \int_0^{x\tan\beta} \left(\frac{x\tan\beta}{h_N} - 1\right)^2 dz + \int_{x_d}^{(h_N+b)/\tan\beta} dx \int_0^{x\tan\alpha} \left(\frac{x\tan\beta}{h_N+b} - 1\right)^2 dz\right]. \quad (5)$$

The energy is minimized, $\partial E(x_d)/\partial x_d=0$, when $x_d(N) = \frac{2N\mathcal{P}(N\mathcal{P}+b)}{(2N\mathcal{P}+b)\tan\beta}$. For $b = \mathcal{P}$, the last expression simplifies to $x_d(N) = \frac{2N\mathcal{P}(N+1)}{(2N+1)\tan\beta}$. The equilibrium distance between two dislocations, $\Delta x_d = x_d(N+1) - x_d(N) = \frac{2\mathcal{P}}{\tan\beta}\frac{(2N\mathcal{P}+b)(\mathcal{P}+b)+2N^2\mathcal{P}^2}{(2N\mathcal{P}+b)(2N\mathcal{P}+2\mathcal{P}+b)}$, for $b = \mathcal{P}$ is then

$$\Delta x_d = \frac{4\mathcal{P}(N+1)^2}{(2N+1)(2N+3)\tan\beta}. \quad (6)$$

The coefficient $c = \frac{4(N+1)^2}{(2N+1)(2N+3)}$ equals 1 with an excellent accuracy for any reasonable $N$. For example, $c = 1.0012$ for $N = 5$ and approaches 1 quickly as $N$ increases. Therefore, the last equation reduces to $\Delta x_d = \mathcal{P}/\tan\beta$. The result supports the conclusion that the observed edge dislocations are of the elementary Burgers vector $b = \mathcal{P}$.

c. Optical retardance around an edge dislocation.

The edge dislocations viewed under a polarizing optical microscope exhibit a variation in interference colors, which are uniquely related to the dilation/compressive stresses, Figure 10. To quantify the effect, we map the optical retardance around the core of an isolated edge dislocation, using a PolScope, Figure 11.

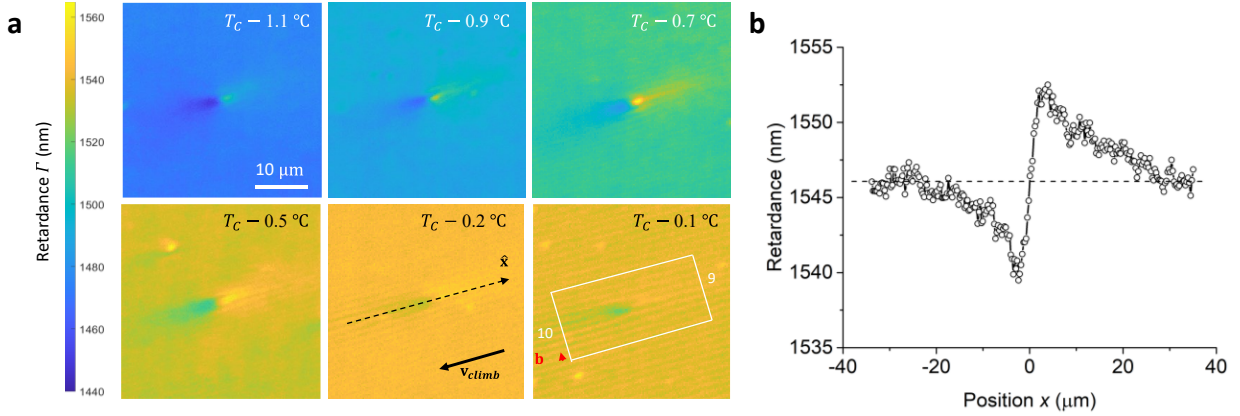

**Figure 11.** Edge dislocations at different temperatures in a $N_{TBF}$ JK203 cell of a thickness $d = (6.6 \pm 0.1)$ μm: a) PolScope (wavelength 535 nm) textures of an edge dislocation $b = \mathcal{P}$ produced by heating of the $N_{TBF}$ towards the transition to $N_F$, which increases $\mathcal{P}$; the edge dislocation climbs along the direction $\mathbf{v}_{climb}$ to remove a pseudolayer; the local retardance is lower in the region with a smaller $\mathcal{P}$ and higher in the area of a larger $\mathcal{P}$. b) Retardance profile of a dislocation at $T_C - 0.2$ °C along the $x$-axis shown in a).

Figure 11a maps the optical retardance of an isolated edge dislocation in a planar cell of a thickness $d = (6.6 \pm 0.1)$ μm upon heating from $T = T_C - 1.1$ °C to $T_C$. The background optical retardance $\Gamma = d\Delta n_{NTBF}$ increases, indicating that the effective $N_{TBF}$ birefringence $\Delta n_{NTBF}$ grows with the temperature. The increase is caused by the decrease of the conical angle $\theta$ since (15)

$$\Delta n_{NTBF}(x,y) = \sqrt{n_e^2 - (n_e^2 - n_o^2)\sin^2\theta} - \sqrt{n_o^2 + \tfrac{1}{2}(n_e^2 - n_o^2)\sin^2\theta} \approx \Delta n_{NF}\left(1 - \tfrac{3}{2}\theta^2\right), \quad (7)$$

where $\Delta n_{NF} = n_e - n_o$, $n_e$ and $n_o$ are the ordinary and extraordinary refractive indices, respectively, of the material that is fully unwound, i.e., in the $N_F$ phase. Measurements in the $N_F$ phase at $T_C + 1.2$ °C yield $\Delta n_{NF} = 0.243$ for the wavelength 532 nm. The optical retardance varies around the dislocation core, Figure 11a. To mitigate the detrimental effect of surface reconstruction, we quantify the variation of retardance around the core at a temperature close to the transition to $N_F$, at $T_C - 0.2$ °C, Figure 11b. The compressed side with an extra layer is of a lower retardance as compared to the background, while the dilated side with a missing layer is of a higher retardance, Figure 11b. The retardance swing is $\delta\Gamma \approx 12$ nm. Therefore, the compressed side exhibits a higher $\theta$ as compared to $\theta$ in the dilated side. The behavior resembles that one of a smectic C, in which a stronger tilt $\theta$ of molecules corresponds to thinner layers.

To estimate the maximum deviation $\pm\delta\bar{\theta}$ at the core from the equilibrium $\theta_0$ far away from the core, we write $\delta\Gamma = \frac{3}{2}d\Delta n_{NF}[(\theta_0 + \delta\bar{\theta})^2 - (\theta_0 - \delta\bar{\theta})^2] = 6d\Delta n_{NF}\theta_0\delta\bar{\theta}$. For $\delta\Gamma = 12$ nm, $\Delta n_{NF} = 0.243$, $d = 6.6$ μm, $\theta_0 = 3°$ at $T_C - 0.2$ °C, Figure 2b, one estimates $\delta\bar{\theta} \approx \pm 1°$.

Polarizing microscopy has already been used to image stresses around edge dislocations in crystalline materials such as silicon carbide (68-70); however, because of low birefringence, one needs to use thick samples (hundreds of micrometers). In the case of $N_{TBF}$, strong birefringence allows one to see the birefringence imprint of dislocations in samples that are only a few micrometers thick.

### d. Core structure of the edge dislocation

The core of edge dislocations of a Burgers vector $b = \mathcal{P}$ in media such as the cholesteric $N^*$ (67, 71, 72) and chiral $N_F^*$ (28) is described within the Kleman-Friedel model (73), in which the core splits into two nonsingular $\lambda$ disclinations of strength +½ and -½ separated by a distance $\mathcal{P}/2$. The helicoidal axis $\mathbf{q}$ rotates by $\pi$ as one circumnavigates around each disclination core. Such a construction is possible since in the cholesteric $N^*$ and the chiral $N_F^*$, $\theta_0 = \pi/2$, and the helicoidal axis is apolar, $\mathbf{q} \equiv -\mathbf{q}$. However, in the $N_{TBF}$, $\theta_0 < \pi/2$ and a non-zero projection of $\mathbf{P}$ onto $\mathbf{q}$ lifts the degeneracy $\mathbf{q} \equiv -\mathbf{q}$. Rotation of $\mathbf{q}$ by $\pi$ produces a domain wall in the polarization field, which hinders the core splitting into disclinations.

The polarization field around the edge dislocation with the core that is not split into +½ and -½ disclinations can be written, neglecting possible variation of the polarization amplitude $P$, as

$$\mathbf{P} = \{P_x, P_y, P_z\} = P\{\sin\theta(x,z)\cos\varphi(x,z),\ \sin\theta(x,z)\sin\varphi(x,z), \cos\theta(x,z)\}, \quad (8)$$

where $\varphi(x,z) = qz + \arctan(z/x)$, as in the model (55) of an edge dislocation in an $N_{TB}$, and $\theta(x,z)$ is the conical tilt. One possible function that qualitatively captures the variation of conical tilt and thus the retardance along the $x$-axis in Figure 11b is

$$\theta(x) = \theta_0 - \delta\theta\frac{x}{L}e^{-\left(\frac{x}{L}\right)^2}, \quad (9)$$

where the second term is a correction to the equilibrium $\theta_0$ that reaches an extremum $\pm 0.43\delta\theta$ at $x/L = \pm 0.7$ and then approaches zero when $|x/L|$ grows beyond 3; $L$ is a characteristic extension of the tilt variation. Figure 12 shows a qualitative scheme of the dislocation core built using **Equation 8** and **9** with $\theta_0 = 0.5$ rad, $\delta\theta = \theta_0/2$, and $L = \mathcal{P}$. It illustrates the experimentally observed higher tilt $\theta$ and smaller $\mathcal{P}$ at $x < 0$ and a lower $\theta$ and larger $\mathcal{P}$ at $x > 0$ in Figure 11.

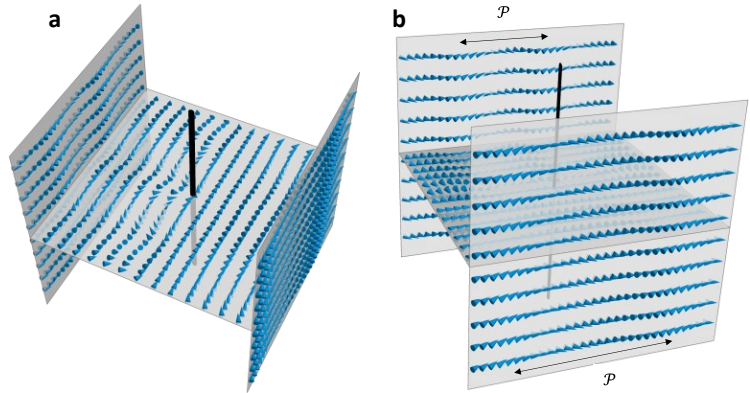

**Figure 12.** Two projections of an elementary edge dislocation $b = \mathcal{P}$ in the N$_{\text{TBF}}$. The thick black rod is the dislocation core.

e. Compressibility of pseudolayers and climb mobility of edge dislocations.

The small value of the compressibility modulus $B \approx 10$ N/m$^2$ explains why the mobility of the edge dislocations is much higher than the mobility of their counterparts in the smectic A phase, as discussed below.

The observed mobility of dislocation is dominated by climb, i.e., by propagation parallel to the pseudolayers and perpendicularly to **q**. In the experiments, we do not observe glide, in which the dislocation moves across the layers and along **q**. The reason is that gliding requires reconstruction of layers at the core, which is hindered by the periodicity of the structure. At the same time, the climb involves movement of molecules within an essentially liquid environment, parallel to the pseudolayers. The prevalence of climb over glide makes the situation similar to the motion of edge dislocations in a smectic A (74-77), N$^*$ (67), and helicoidal magnets (78). The mobility of dislocations is defined as the ratio $M = \frac{v}{\sigma_{zz}}$, where $v$ is the (climb) velocity, and $\sigma_{zz} = B\frac{\partial u}{\partial z} \approx B\frac{\partial \mathcal{P}}{\partial T}\frac{\Delta T}{\mathcal{P}}$. The average experimental velocities are about 30 µm/s in the high-temperature range of the N$_{TBF}$ phase during heating, where the thermal expansion coefficient is $\frac{\partial \mathcal{P}}{\partial T}\frac{1}{\mathcal{P}} \approx \frac{0.3}{1.2K} \approx 0.25/K$, while the rate of temperature change is 0.0167 K/s. Therefore, the mobility can be estimated as $M = 7 \times 10^{-4}\frac{m^2 s}{kg}$ during heating in the high-temperature end. A similar estimate $M = 6 \times 10^{-4}\frac{m^2 s}{kg}$ is produced in the low-temperature range upon heating, where the velocities are down to 10 µm/s, while the thermal expansion is $0.1/K$. For the cooling regime, at the low-temperature range of the N$_{TBF}$ phase, where the velocities are on average 7 µm/s, and the thermal expansion is $0.055/K$, the estimate is practically the same, $M = 8 \times 10^{-4}\frac{m^2 s}{kg}$. These mobilities are substantially higher than the mobilities measured previously for the smectic A, $M = 2 \times 10^{-8}\frac{m^2 s}{kg}$ in Ref.(75), and $4.2 \times 10^{-8}\frac{m^2 s}{kg}$ in Ref. (77), which is naturally related to the smallness of $B$ in the N$_{TBF}$ phase. Note also that another factor contributing to the rapid climb of edge dislocations is the absence of a composite core with +1/2 and -1/2 disclinations, as in the case of N$^*$ dislocations. In the latter case, the pseudolayers between the two disclinations are oriented perpendicularly to the $x$-axis and climb is hindered by the permeation phenomenon.

## IV. CONCLUSION

The ferroelectric twist-bend nematic exhibits several remarkable structural and optical properties, including temperature- and field-controlled periodicity. Our study of N$_{TBF}$ cells with controlled surface anchoring reveal essential facets of the macroscopic structural organization of this mesophase.

1. Alignment of the N$_{TBF}$ in sandwich-type cells keeps the vector **q** of the of twist-bend heliconical axis predominantly in the plane of the cell, which results in fingerprint textures such as the one in Figure 4a. Such an alignment is dictated by electrostatics, which prevents **q** and the polarization **P** from piercing the interface with the glass plates and depositing bound surface charges.

2. The strong temperature dependence of the pitch $\mathcal{P}$ and surface interactions make the surface periodicity of N$_{TBF}$ different from the bulk value of $\mathcal{P}$. The frustration is resolved by a tilt of pseudolayers in thin cells, $d < d_{ch}$, and by chevrons in thick cells, $d > d_{ch}$; here $d_{ch} = 7 - 9$ µm is the critical cell thickness of the chevron formation. In these textures, **q** tilts up and down along the normal to the cell, Figure 7b,c. With the measured temperature dependence of $\mathcal{P}$ and the expected stresses caused by the periodicity mismatch between the surface and bulk, these data are in good agreement with the Limat-Prost model of tilted and chevron structures.

3. Unlike the fingerprint textures of cholesteric liquid crystals, in which the twist axis is perpendicular to the surface easy axis (such as the rubbing direction **R**), the projection of **q** onto the bounding plates is not collinear to the surface-imposed rubbing direction **R**. We associate this behavior with the surface reconstruction which reduces the normal component of polarization and the bound charge at the surface, Figure 4c. Surface reconstruction results in a mismatch of surface and bulk periodicities and also introduces optical activity of textures along the normal to the cell.

4. The strong temperature dependence of $\mathcal{P}$ results in the appearance of numerous edge dislocations that gather into tilt grain boundaries whenever the sample is cooled or heated. The profile of pseudolayers around the dislocation cores allows us to estimate the elastic penetration length $\lambda = \sqrt{K/B}$ of the $N_{TBF}$ as being on the order of $\mathcal{P}$. Unlike the edge dislocations in paraelectric and ferroelectric cholesterics, the cores of edge dislocations do not split into ½ and -½ disclinations because of the polar character of the **q** axis, $\mathbf{q} \neq -\mathbf{q}$. Stresses around the dislocation cores result in variations in the optical retardance associated with different conical tilt angles. Near the transition to the $N_F$, the conical tilt angle is larger in the compressed region of the core and smaller in the dilated region. Since the core does not split into disclinations, the edge dislocations show high climb mobility.

The uncovered structural behavior of the $N_{TBF}$ might be helpful in the development of optical elements, such as diffraction gratings and structural colors, which an electric field or temperature can tune. The observed high mobility of edge dislocations is of a potential advantage in speeding up the adjustment of pitch.

**ACKNOWLEDGEMENTS**

The work was supported by NSF grant DMR-2341830 (ODL) and NCN (Poland) project OPUS 2024/53/B/ST5/03275 (PK, DP, and EG). NV acknowledges the support of the Slovenian Research and Innovation Agency (ARIS), through the research core funding Program No. P1-0055. ODL is thankful to the University of Warsaw for the support during the sabbatical leave from Kent State University, under IDUB II.2.1 programm, WNE II.2.1/02/2024.

**Data Availability Statement**

All data needed to evaluate the conclusions in the paper are present in the article and in the supplementary figures.

**Conflict of Interests**

The authors declare no confliuct of interest.

# Helix alignment, chevrons, and edge dislocations in twist-bend ferroelectric nematics

Bijaya Basnet[1,2], Priyanka Kumari[1,2], Sathyanarayana Paladugu[1], Damian Pociecha[3], Jakub Karcz[4], Przemysław Kula[4], Nataša Vaupotič[5,6], Ewa Górecka[3], and Oleg D Lavrentovich [1,2,3,7*]

[1]*Advanced Materials and Liquid Crystal Institute, Kent State University, Kent, OH 44242, USA*
[2]*Materials Science Graduate Program, Kent State University, Kent, OH 44242, USA*
[3]*Faculty of Chemistry, University of Warsaw, Zwirki i Wigury 101, Warsaw 02-089, Poland*
[4]*Faculty of Advanced Technology and Chemistry, Military University of Technology, Warsaw, Poland*
[5]*Jozef Stefan Institute, Jamova 39, 1000 Ljubljana, Slovenia*
[6]*Department of Physics, Faculty of Natural Sciences and Mathematics, University of Maribor, Koroška 160, 2000 Maribor, Slovenia*
[7] *Department of Physics, Kent State University, Kent, OH 44242, USA*

*Authors for correspondence: e-mails: olavrent@kent.edu

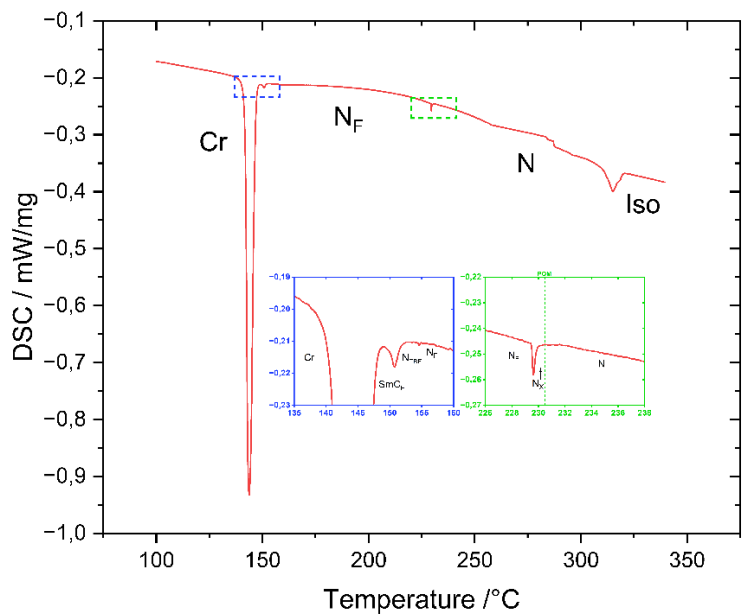

Figure S1: Differential Scanning Calorimetry (DSC) for JK203. DSC thermogram recorded on heating (10 K/min).

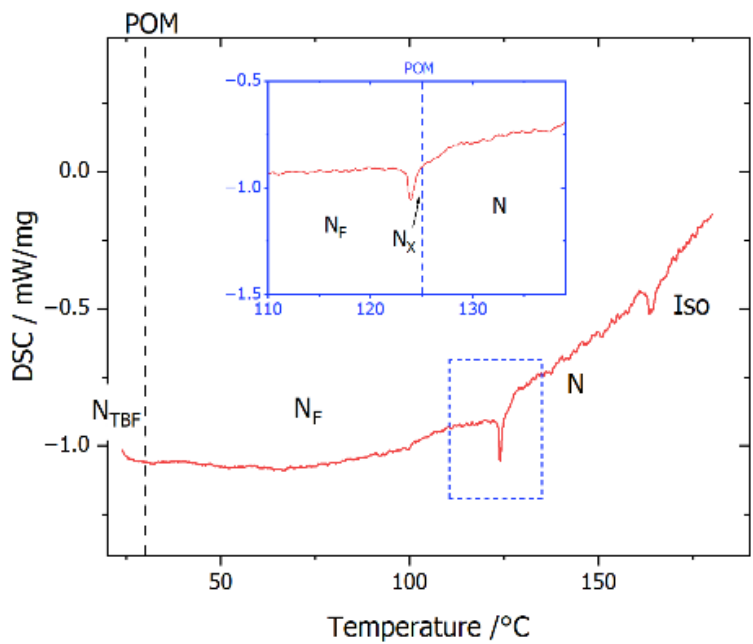

Figure S2: DSC for NTBF005. DSC thermogram recorded on heating (10 K/min). Dashed lines show $N_{TBF}$-$N_F$ and $N_X$-N transition temperatures determined from optical studies.

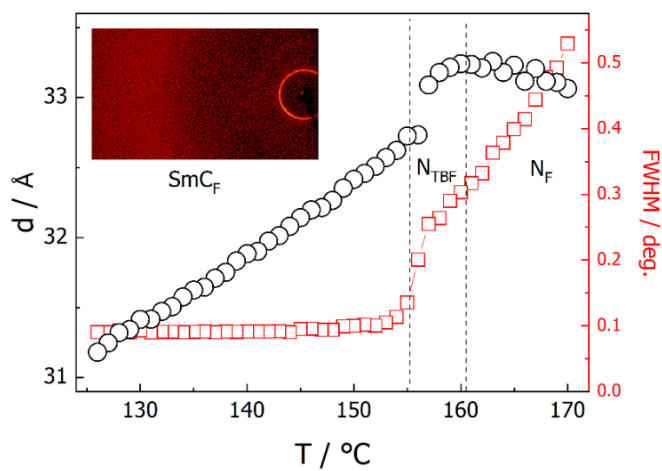

Figure S3: X-ray diffraction (XRD) for JK203. Layer spacing in the smectic phase and local periodicity detected in nematic phases of JK203 by small angle X-ray diffraction (open circles); full width at half maximum (FWHM) of the diffraction signal (red squares) showing gradual development of positional order in nematic phases, which in smectic phase becomes truly long-range (the width of the XRD signal related to density periodicity is machine resolution limited). The inset shows a 2D XRD pattern recorded in wide angle range in the $SmC_F$ phase.

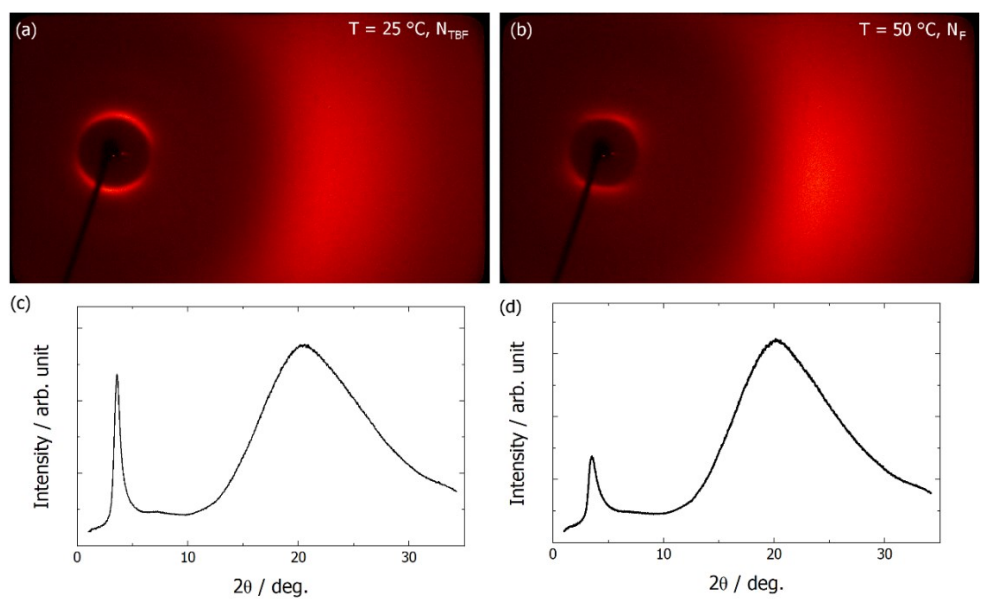

Figure S4: XRD for NTBF005. 2D XRD patterns recorded in wide angle range in (a) $N_{TBF}$ and (b) $N_F$ phases. (c) and (d) Diffracted intensity vs. diffraction angle obtained by integration over the azimuthal angle of the patterns presented in (a) and (b), respectively. In both phases all diffraction signals are broadened with respect to instrumental resolution, reflecting lack of long-range positional order of molecules.

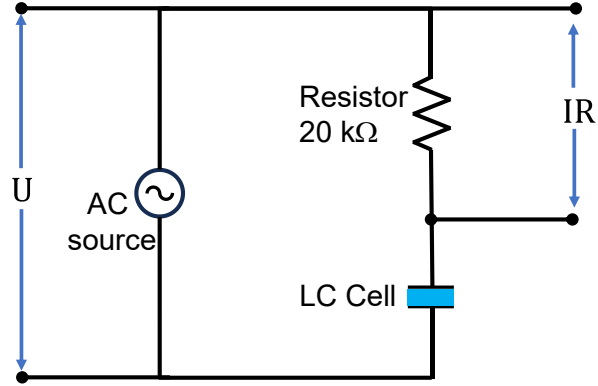

Figure S5: Electric circuit to measure the polarization density from the polarization reversal produced by an in-plane electric field. The ITO-glass substrates of 100 Ω/□ resistance were patterned lithographically to produce 2 ITO electrode stripes (2 mm × 14 mm) on the 11 mm × 16 mm glass area. The electrodes are separated by a 1 mm gap. The ITO-glass plates are covered with a polyimide layer which is rubbed to set the surface alignment collinear with the in-plane electric field. An in-plane ac electric field of a triangular waveform, frequency 50 Hz, and peak-to-peak voltage of 240 V was applied using a Siglent SDG1032X waveform generator and an amplifier (Krohn-Hite corporation). Cells of a gap thickness $d = (20.0 \pm 0.1)$ μm are used. The electric current through a 20 k Ω resistor is measured with an oscilloscope Tektronix TDS 2014 (sampling rate 1GSa/s). When the polarity of the triangular wave is reversed, the net polarization charge produced by the polarization reversal is $Q = 2PA$, where $A = 20$ μm × 1.4 cm is the cross-sectional area of the liquid crystal. The polarization density is calculated as $P = \frac{\int I(t)\, dt}{2A}$ where $\int I(t)\, dt$ is the area under the polarization current curve. A similar set-up is used to verify the polarization alignment with respect to the rubbing direction; in this case, thin cells, $d = (2.3 \pm 0.1)$ μm, are used to avoid twist deformations.

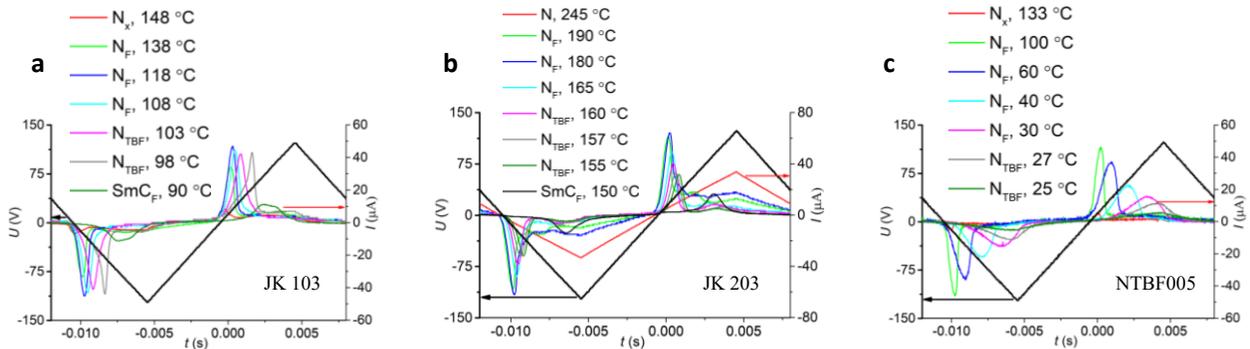

Figure S6: Polarization density measurements for a) JK103, b) JK203, and c) NTBF005. The time dependence of applied voltage (triangular waveform, $f = 50$ Hz and peak-to-peak voltage 240 V) and electric current. Thicker black solid line represents the applied voltage whereas thinner solid lines represent the electric currents at different temperatures. $d = (20.0 \pm 0.1)$ μm; $A = 20$ μm × 1.4 cm.

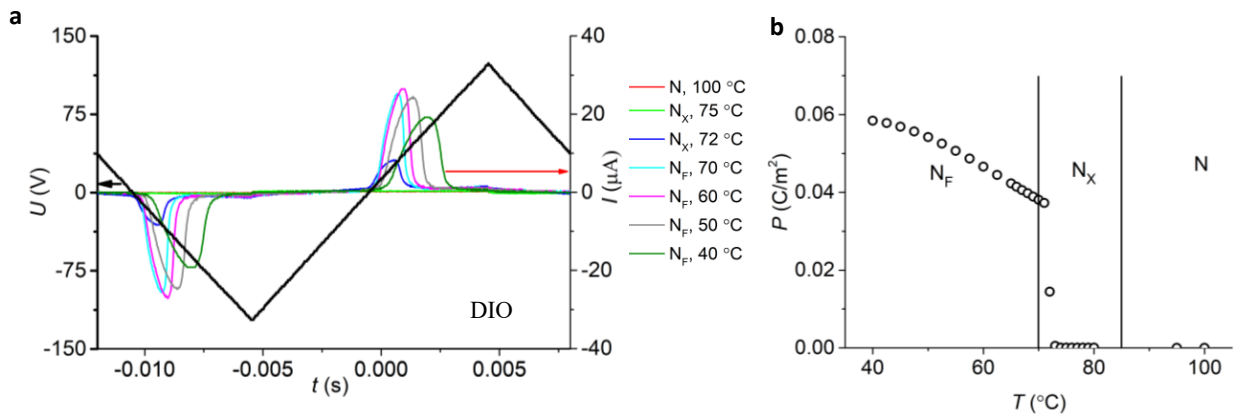

Figure S7: Polarization density of DIO as a function of temperature: a) time dependence of applied voltage (triangular waveform, $f = 50$ Hz and peak-to-peak voltage 240 V) and electric current. Thicker black solid line represents the applied voltage whereas thinner solid lines represent the electric currents at different temperatures; b) The polarization density as a function of temperature in the $N_F$ phase. $d = (20.0 \pm 0.1)$ μm; $A = 20$ μm × 1.4 cm.

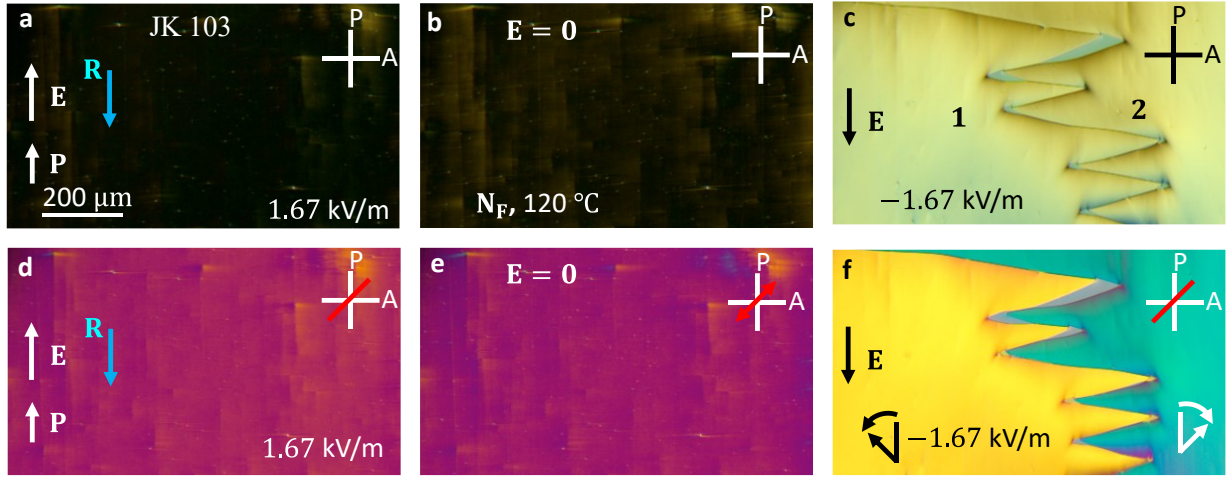

Figure S8: Polarizing microscope textures of a uniform polarization alignment in thin planar $N_F$ cell of JK103. a-c) A direct current (dc) electric field antiparallel to the rubbing direction **R** (**E**↑**R**↓) does not change the interference color while application of the field along **R** (**E**↓**R**↓) rotates **P** counterclockwise and clockwise, producing domains 1 and 2, respectively; d-f) the same, observations with a full-wavelength 550 nm optical compensator. $d = (2.3 \pm 0.1)$ µm; 120 °C. The electric field response illustrates that **P** is antiparallel to the buffing direction **R**.